\documentclass[letterpaper,12pt]{article}   
\usepackage{osajnl2} 
\usepackage[draft]{hyperref} 

\newcommand{\ud}{\,\mathrm{d}}
\newcommand{\aap}{A\&A}
\begin{document}

\title{
      Precision in ground based solar polarimetry: \\
      Simulating the role of adaptive optics
      }


\author{Nagaraju Krishnappa and Alex Feller}
\address{Max Planck Institute for Solar System Research, Katlenburg-Lindau, \\ 37191, Germany}
\address{$^*$Corresponding author: krishnappa@mps.mpg.de}

\begin{abstract}
Accurate measurement of polarization in spectral lines is important for 
the reliable inference of magnetic fields on the Sun. 
For ground based observations, polarimetric precision is severely limited
by the presence of Earth's atmosphere.
Atmospheric turbulence (seeing) produces signal fluctuations 
which combined with the non-simultaneous nature of the measurement 
process cause intermixing of the Stokes parameters
known as seeing induced polarization cross-talk.
Previous analysis of this effect \cite{judge04} suggests that cross-talk 
is reduced not only with increase in modulation frequency but also by 
compensating the seeing induced image aberrations by an Adaptive Optics (AO) 
system. 
However, in those studies the effect of higher order image aberrations than 
those corrected by the AO system was not taken into account. 
We present in this paper an analysis of seeing induced cross-talk in the 
presence of higher order image aberrations through numerical simulation. 
In this analysis we find that the amount of cross-talk among Stokes parameters 
is practically independent of the degree of image aberration corrected by an 
AO system.
However, higher order AO corrections increase the signal-to-noise ratio
by reducing the seeing caused image smearing.
Further we find, in agreement with the earlier results, that cross-talk 
is reduced considerably by increasing the modulation frequency.
\end{abstract}


\maketitle 

\section{Introduction}
Polarimetry has become an essential part of back-end instrumentation of modern
solar telescopes. High precision polarization measurements are essential
to infer the magnetic field information on the Sun. 
For ground based observations, the presence of atmospheric turbulence, commonly 
known as seeing, is one of the major limiting factors in achieving high 
precision in polarization measurements.
A polarization state, which is the result of the phase difference between
the orthogonal components of the electric field vector in the light beam, 
is not directly detectable by photo-detectors.
Hence the polarization information is converted into 
intensity by modulating the light in a known way. Depending on the
modulation scheme several intensity measurements are required, 
with a minimum of four, in order to recover the complete polarization 
information (i.e. the four Stokes parameters).
Often these measurements are obtained at successive intervals making
the polarization measurement a non simultaneous process. Rapid changes
in seeing conditions can therefore produce spurious polarization signals in 
polarimetric data.

A Fourier transformation based formalism has been developed by 
Lites \cite{lites87} to analyze seeing induced cross-talk. 
He considered to analyze the cross-talk terms caused only by the seeing induced 
image motion (tip-tilt terms).
He finds that the cross-talk terms can be reduced
considerably by increasing the modulation frequency.
This initial work by Lites was revised and extended by
Judge et al. \cite{judge04} to study the effect of residual image motion after
a partial correction of the tip-tilt terms by an AO system. 
In addition to the rotating-waveplate modulation scheme, originally studied
by Lites, they also consider a discrete modulation scheme based on
liquid crystal variable modulators.
The studies of Judge et al. suggest that with an increase in the
degree of image aberration corrections,  one can achieve a reduction in 
seeing induced cross-talk. 
This means that if a polarimeter is used in conjunction with an AO system,
a significant reduction in seeing induced polarization is expected. 
However, in their work the effect of higher order aberration terms is not 
taken into account.

Initial results on the effect of higher order aberration terms have been
obtained by Nagaraju et al. \cite{nagaraju11} through a numerical simulation.
In that study the analysis is done for a 4~m diameter telescope  with
only a limited number of higher order terms compensated.
However, our aim is to assess seeing induced cross-talk in the case when AO
performance is close to the diffraction limit.
To accomplish the diffraction limited performance of AO, the number
 of terms compensated is much beyond the practical limit because of
the asymptotic way the diffraction limit is achieved 
(cf.  Section \ref{sec:imaging}).
In order to characterize the phase screens and images after AO corrections
close to the diffraction limit, in a statistically consistent way, for the 
case of a 4~m aperture telescope the simulation becomes computationally expensive.
Hence we opted to carry out the simulation for a 1~m aperture telescope in this 
work with AO performance close to the diffraction limit. The computational
resources required are multifold less for a 1~m than for a 4~m aperture telescope.
Another reason for choosing 1~m aperture telescope is that the results 
presented in this paper could possibly be verified through
actual measurements using one of the currently operating 1~m class telescopes.

\section{Numerical Simulation of Imaging Through Earth's Atmosphere with AO Correction}
Turbulence in Earth's atmosphere produces inhomogeneity in the refractive
index. This causes fluctuations in the phase of the light wave
across the telescope aperture, known as phase screen.
As a result, when the imaging is done through the atmosphere, 
image quality is severely degraded.
The amount of image degradation increases with increase in ratio of the
aperture diameter ($D$) to the Fried parameter $(r_0)$.
In order to achieve higher image quality, in other words to achieve
a spatial resolution closer to the diffraction limit, most 
modern solar telescopes are equipped with an AO system. 
An AO system corrects for
a number of lowest order Zernike or Karhunen-Lo{\'e}ve modes (see e. g., 
Roddier \cite{roddier99} and references therein).
Our aim is to understand seeing induced cross-talk
among Stokes parameters in the context of an AO system working with a telescope.
The steps involved in the numerical simulation are as follows.
Time dependent phase screens above the telescope of 1~m diameter are
generated. Then synthetic Stokes images are convolved with the time dependent
point spread functions (PSFs) corresponding to the phase screens. 
The resulting time series of convolved Stokes images is then
polarimetrically analyzed using a chosen modulation scheme.
In the following we discuss these steps in detail.

\subsection{Phase Screen Simulation}
We use the power spectral method to generate phase screens 
(McGlamery \cite{mcglamery76}). 
In this section we briefly describe the method. More details can be 
found in \cite{mcglamery76, lane92, glindemann93}.  

The Kolmogorov power spectrum is given by
\begin{equation}
\label{eq:powspec}
\Phi(\vec{k}) = 0.023 r_0^{-5/3} |\vec{k}|^{-11/3},
\end{equation}
where $r_0$ is Fried parameter in meters and $\vec{k}$ is 2-D spatial
frequency vector.
The above expression can be written in discrete form \cite{lane92} as
\begin{equation}
\label{eq:powspecdiscrt}
\Phi(k,l) = 0.023 \left(\frac{NS}{r_0}\right)^{5/3} \left(\sqrt{k^2+l^2}\right)^{-11/3},
\end{equation}
with $k,l$-integer array indices, $N^2-$size of the square array and $S-$pupil sampling
in meters per pixel.

The magnitude of the Fourier spectrum of phase screen is given by
\begin{equation}
|\hat{\varphi}(k,l)| = \sqrt{0.023} \left(\frac{NS}{r_0}\right)^{5/6} \left(\sqrt{k^2+l^2}\right)^{-11/6}.
\label{eq:magFS}
\end{equation}
The Fourier transform of the product of magnitude of the Fourier spectrum in 
Eq. \ref{eq:magFS} and a 2-D array of Gaussian distributed random numbers with 
variance one yields a realization of a phase screen, at a given point in time. 

   \begin{figure}
   \includegraphics[width=0.70\textwidth,height=0.35\textheight]{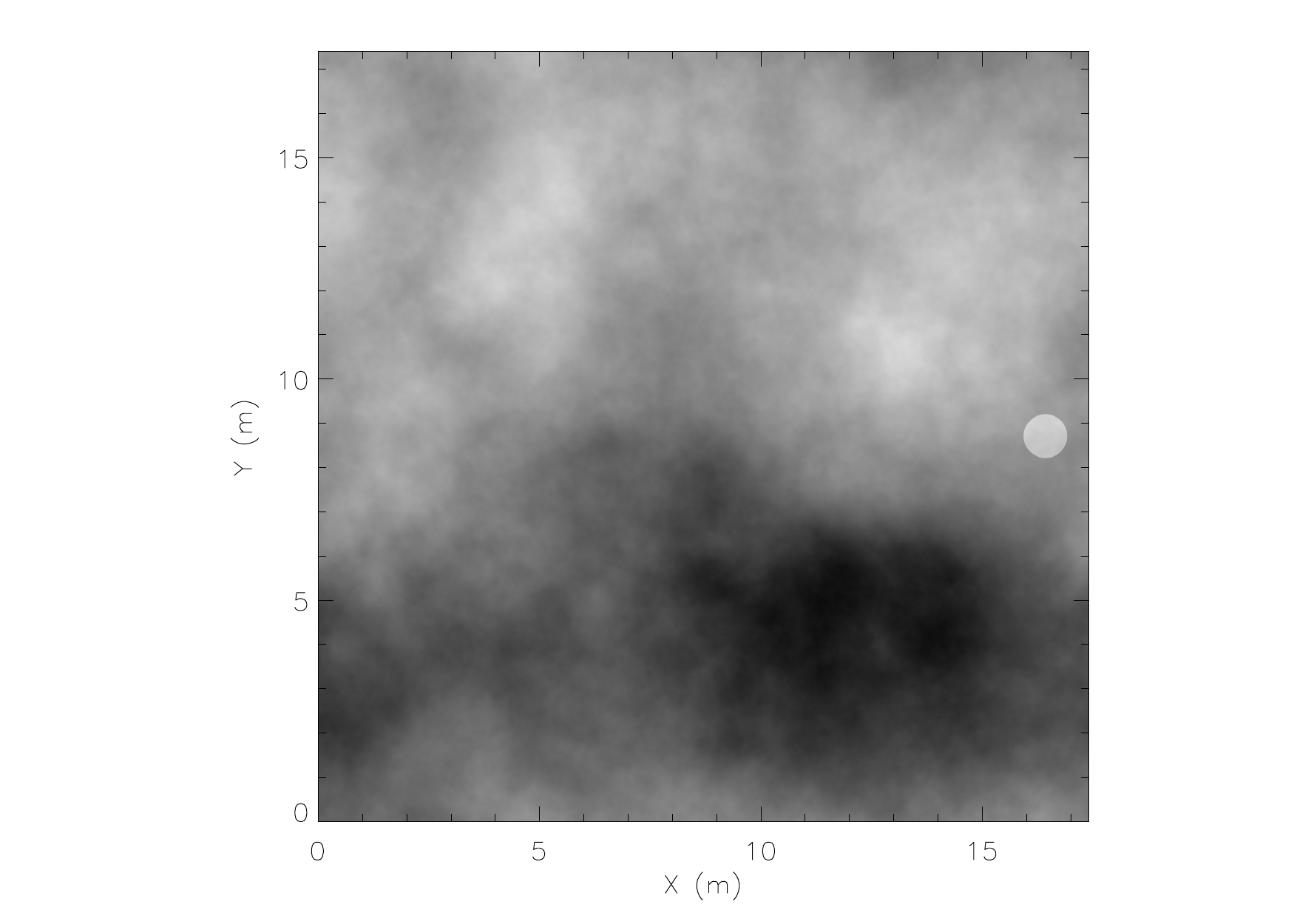}
   \caption{
   A sample phase screen with an enhanced circular region on the right side 
   representing a telescope aperture of 1~m. The grey scale is $\pm 19$ waves.
   The simulation parameters are compiled in Table 1.
   }
   \label{fig:phasescreen}
   \end{figure}

Time evolution of phase screens above the telescope aperture is assumed to
be produced by wind driven propagation of frozen in turbulence 
\cite{roddier93}.
In the simulation this is achieved by generating a large phase screen several 
times bigger than the aperture size and shifting it across the telescope 
aperture with constant speed.
Wind effect is introduced by adding a phase in the Fourier
spectrum i.e.
\begin{equation}
\hat{\varphi}_{wind}(k,l) = \hat{\varphi}_{old}(k,l) e^{2\pi i(kN_k+lN_l)/N}.
\end{equation}
Where $N_{k,l}$ are the number of pixels to be shifted.
A sample large phase screen is shown in Fig. \ref{fig:phasescreen} with
an enhanced circular region representing a 1~m telescope aperture.
The simulation parameters are compiled in Table 1.
The wind propagation direction is assumed to be in positive $X$-direction.
The simulation parameters are the same throughout this paper
unless otherwise specified.

\begin{table}
\label{tab:simpar}
\begin{center}
\begin{tabular}{|c|cc|}
\hline
Simulation & Value&\\
parameter & &\\
\hline
$N$ & 2048&\\
$S$ & 0.85& cm/pixel \\
$r_0$ & 10.0& cm \\
$v$ & 10.0& m/s \\
$\Delta t$ & 62.5&$\mu$s\\
\hline
\end{tabular}
\caption{
         Simulation parameters. The parameter $v$ is the wind speed and 
         $\Delta t$ is the sampling interval. See Eq. \ref{eq:powspec} or 
         Eq. \ref{eq:powspecdiscrt} for a description of the other parameters.
         }
\end{center}
\end{table}

We would like to note here that the phase screens clipped to 1~m aperture do 
not exactly show Kolmogorov power law behavior. 
The decrease in their power with frequency is slower than the Kolmogorov power. 
As pointed out by \cite{lane92}, this is due to the fact that 
the convolution of the Kolmogorov power spectrum with the power spectrum of 
the telescope aperture results in a power law between $k^{-2}$ and $k^{-11/3}$. 
Secondly, for the large phase screen the piston term (zero frequency) is set 
to zero. Since the clipping of the phase screens changes the zero
frequency, it is reset to zero as well, for all phase screens. 
Further, we did not consider to include subharmonics in our simulation, as
it is usually done to account for underestimation of lower order terms
(mainly the tip/tilt terms) 
when the power spectral method is used to simulate phase screens \cite{lane92}. 
This is because our interest in this work is to study the effect of higher 
order terms.

\subsection{Imaging and Correction of Aberrations by an AO System}
\label{sec:imaging}
\begin{figure}
   \includegraphics[width=0.76\textwidth,height=0.33\textheight]{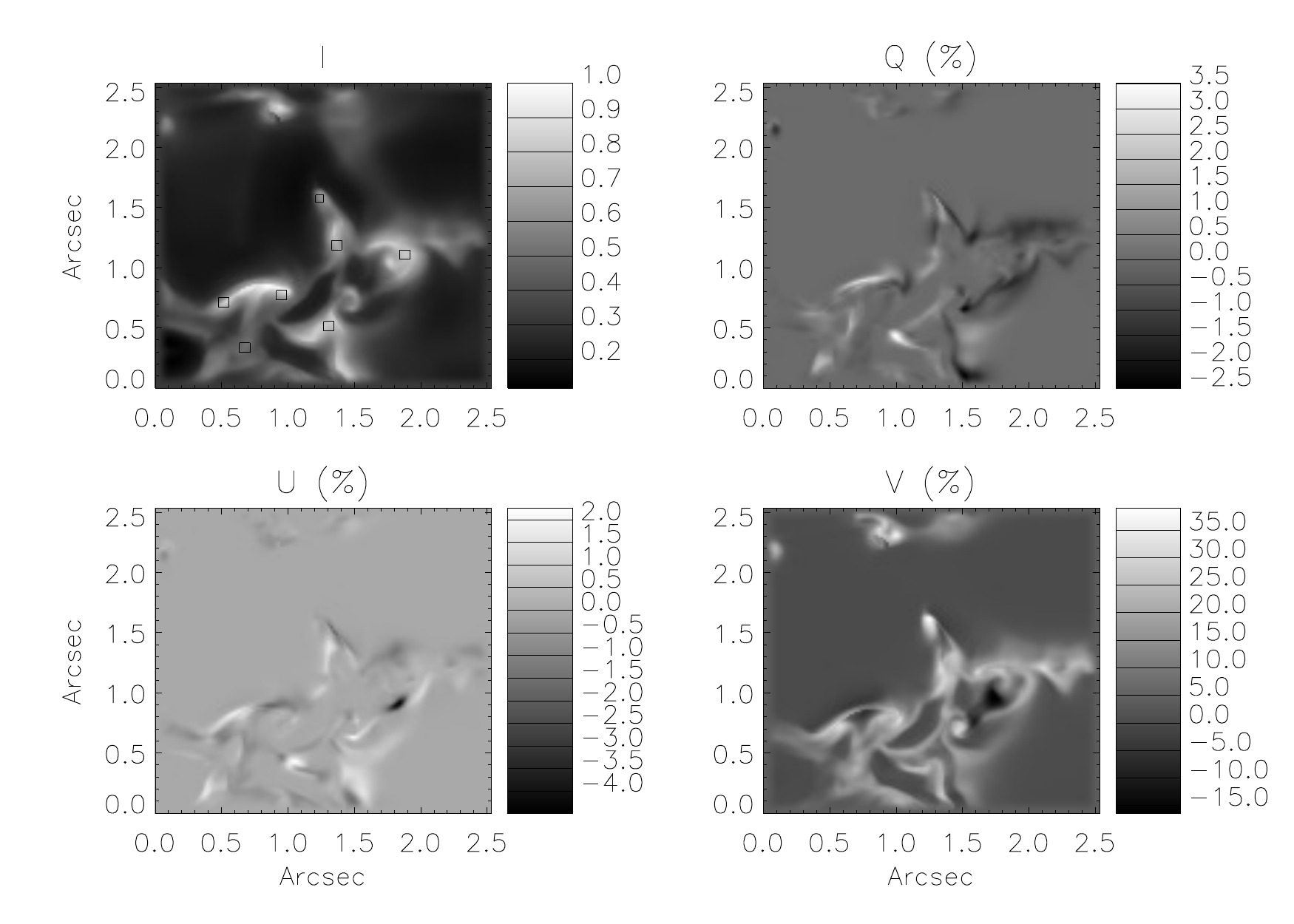}
   \caption{
   Synthetic Stokes images representing a plage region in the wing (+24~m\AA) of
   the Fe {\sc i} line at 6302~\AA. The rectangular boxes are the
   regions considered to analyze cross-talk terms. The Stokes images are
   normalized to the maximum value of Stokes $I$ in the field-of-view.
   }
   \label{fig:stoksynth}
   \end{figure}

Seeing induces cross-talk among Stokes parameters through randomly spreading
information, temporally and spatially, of one detector pixel to the other. 
Hence the amount of cross-talk induced also depends on the structuring of 
Stokes intensities on the Sun. 
This fact is taken into account in Lites formalism by taking the 
cross-talk term proportional to the Stokes intensity gradient.
Hence the comprehensive assessment of image formation affected by the
seeing induced optical aberrations and after AO correction is very 
important to understand seeing induced cross-talk. 
Imaging through Earth's atmosphere by a telescope is simulated by convolving
the synthetic Stokes images with the PSFs derived from the corresponding
phase screens above the telescope aperture. 
We consider synthetic Stokes images representing a plage region on the Sun 
as input (Fig. \ref{fig:stoksynth}). 
The images have been synthesized by N. Vitas applying the LILIA radiative
transfer code by Socas-Navarro \cite{socas-navarro01} on a 
MURaM (MPS/University of Chicago Radiative MHD) cube 
\cite{voegler03,voegler04,voegler05}.
The image scaling of the synthetic images is 0.0198~$''$/pixel.
The synthetic images are selected at 24~m\AA\, in the red wing of the 
Fe~{\sc i} line at 6302~\AA.
The detector is assumed to have
$128\times128$ pixels. 
The boxes in this figure represent the regions considered for cross-talk
analysis each covers the area of about $0.099'' \times 0.099''$.

To simulate the correction due to AO a number of lowest order Zernike terms
are compensated in the simulated phase screens. 
The Zernike coefficients are estimated as
\begin{equation}
a_j(t) = \int\!\!\!\int{\varphi(\vec{r},t)Z_j(\vec{r}) \ud \vec{r}},
\end{equation}
where $\vec{r}$ is a two dimensional spatial vector, $\varphi(\vec{r},t)$ is a 
realization of the phase screen above the telescope at an instant $t$ and
$Z_j(\vec{r})$ is the Zernike polynomial of order $j$. The Zernike order number
follows the convention used in Noll \cite{noll76}.
In order to carry out the integration we have used the Gauss-cubature 
method \cite{stroud71} which provides high numerical accuracy in 
the estimation of $a_j$. 
The original code to carry out 2-D integration over a disk written in C$^{++}$ 
by Holoborodko \cite{holoborodko} has been converted into Interactive Data Language (IDL) and is used in this work.
The residual phase screen, after correcting for $J$ number of lowest order 
Zernike terms, compensated is given by
\begin{equation}
\label{eq:resps}
\varphi_J(\vec{r},t) = \varphi(\vec{r},t) - \sum_{j=1}^{J} a_j(t) Z_j(\vec{r}).
\end{equation}

\begin{figure}
   \includegraphics[width=0.99\textwidth,height=0.4\textheight]{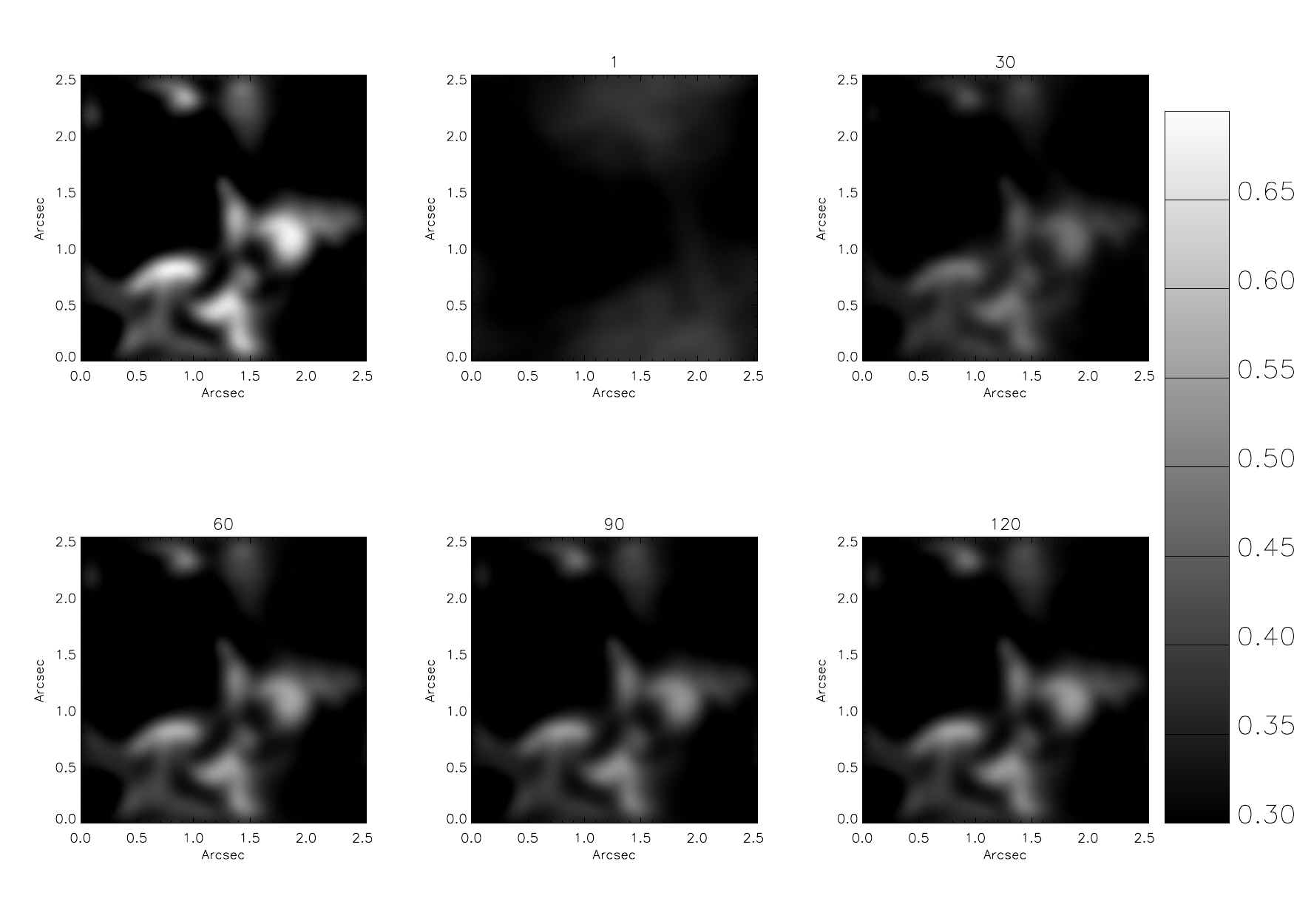}
   \caption{
   Convolved Stokes $I$ images with the phase screens of diameter 1~m. 
   The top left panel contains the diffraction limited image of a 1~m aperture
   telescope.
   The number on each other image indicates the number of lowest order Zernike 
   terms compensated in the phase screen (cf. Eq. \ref{eq:resps}).
   All the images are normalized to the maximum value of input
   Stokes $I$ in the field-of-view.
   }
   \label{fig:intconv1m}
   \end{figure}

In the following we discuss the performance of an ideal AO system.
This means that the AO system fully compensates for $J$ number of lowest order
Zernike terms.
For demonstration purpose we discuss in detail the
image formation through Earth's atmosphere and AO correction in the case of a 
Stokes $I$ image (Fig. \ref{fig:intconv1m}). 
The conclusions drawn from this analysis apply to the other Stokes images as well.
The synthetic Stokes $I$ image is convolved with the instantaneous
PSFs corresponding to the residual phase screens of Eq. \ref{eq:resps}.
Before convolution, the synthetic Stokes images are subjected to apodization in
order to avoid the ringing effect which is produced in the Fourier spectrum due
to the sharp edges present in the synthetic Stokes images.

\begin{figure}
   \includegraphics[width=0.79\textwidth,height=0.35\textheight]{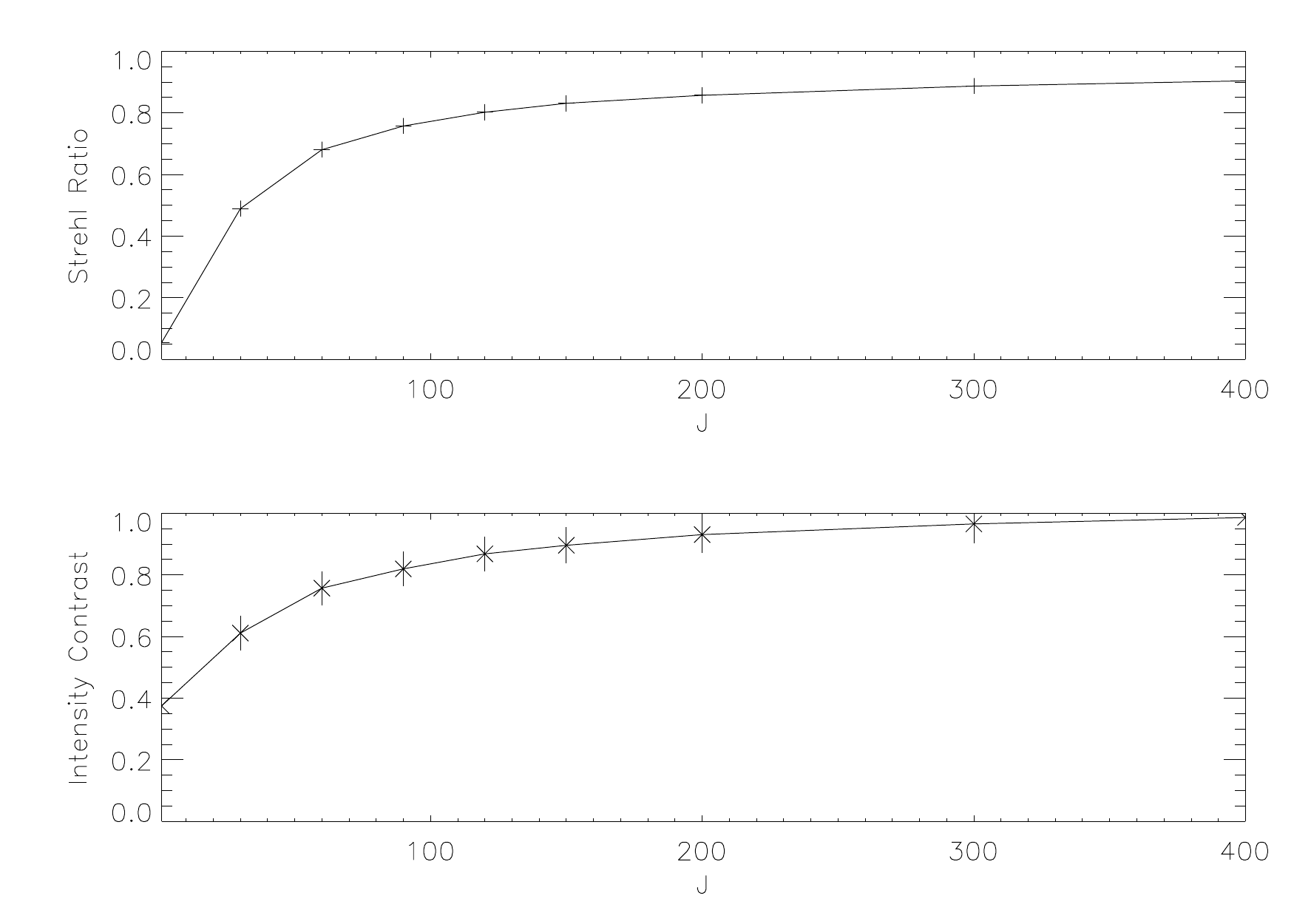}
   \caption{
   Plots of Strehl ratio (top panel) and intensity contrast (lower panel) 
   versus the number of lower order Zernike terms compensated.
   }
   \label{fig:strehl}
   \end{figure}

Increase in the number of Zernike terms compensated brings the convolved images 
closer to the diffraction limited image. 
For the Fried parameter considered in the simulation 
(cf. Table 1), with 30 terms 
compensated the structures in the convolved image already look very close to 
the structures in the diffraction limited image, although the contrast is 
still considerably lower (60\% of the diffraction limited value).
The Strehl-ratio (SR) of the corresponding PSF is about 0.48. 
With increase in $J$ terms compensated beyond 60 there is hardly any 
visually significant change in the convolved images.
However, the SR  and intensity contrast show asymptotic behavior with
respect to $J$ (Fig. \ref{fig:strehl}).
The values of SR smaller than unity and the reduced intensity contrast are
 due to the presence of residual higher order aberrations in the 
AO corrected phase screens. 
Even with as many as 400 terms compensated the SR is only about 0.9 and the
intensity contrast is about 99\% of the diffraction limited value.
In practice, the maximum number of terms compensated is about equal to the 
number of actuators used in the AO system which is in the order 
of $(D/r_0)^2$ \cite{northcott99}. 
In the case considered for simulation it is practical to correct up to about 
100 terms. 
This implies that even with an ideal AO system, which corrects for 
the possible maximum number of aberration terms, the residual phase errors are 
still significant.
Since these residual phase errors change rapidly with time, they play an 
important role in producing polarization cross-talk.

\subsection{Polarization Modulation Schemes}
We consider to study two modulation schemes. One is based on a continuously 
rotating waveplate \cite{lites87} and another on two ferro-electric liquid 
crystal cells (FLCs) in combination with two static retarders \cite{keller03}.
These two schemes provide a direct comparison between continuous and discrete
modulation schemes. 

Polarimetric measurements are simulated with the polarimeter operating in a 
single- or a dual-beam setup. 
In the case of dual beam setup the input beam is split into two orthogonally
polarized beams using a beam splitter and measured simultaneously. 
In the case of singlebeam setup orthogonally polarized beams are measured 
sequentially.

\subsubsection{Rotating Waveplate Modulation Scheme (scheme 1)}
In this scheme the light beam is modulated by a continuously rotating waveplate 
followed by a polarization analyzer.
The polarization analyzer is a simple linear polarizer in the case
of single-beam setup and it is a polarizing beam splitter in dual-beam setup.
We adopt the same detection scheme as described in Lites \cite{lites87}, 
the one in which modulated intensities are sampled at eight equally spaced 
intervals during one half rotation of the waveplate.
This implies the exposure time of $T/16$, where $T$ is the rotation period of 
the retarder. The stokes parameters are obtained by evaluating linear
combinations of these modulated intensities. 
In our simulation we assume a retardance value of 150$^o$ for the waveplate.

\subsubsection{FLC Modulation Scheme (scheme 2)}
In this scheme two FLCs, each followed by a static waveplate, are used for
polarization modulation, and a polarizer for polarization analysis.
The FLCs are assumed to be half-wave plates and static retarders to be
quarter-wave plates.
Polarization modulation is done at discrete intervals by switching the
fast axes of the FLCs between their bistable states.
We adopt the scheme described in \cite{keller03} in which
light is modulated with four different combinations of FLC fast axis
 position angles.
This implies four intensity measurements with the exposure time of 
$T/4$, where $T$ is the modulation period.
The Stokes parameters are obtained by taking a linear combination of
these four modulated intensities.

\section{Results and Discussions}

The input Stokes images are convolved with the time dependent PSFs 
corresponding to the phase screens after correcting for $J$ Zernike terms.
These convolved images are then subjected to polarization analysis
using the chosen modulation scheme. 
The cross-talk among Stokes parameters is estimated by inputting 
a pure Stokes parameter, either Stokes $I$, $Q$, $U$ or $V$ is input
at a time.

The cross-talk from Stokes parameter $i$ to $r$ is estimated as
\begin{equation}
\label{eq:Cri}
C_{ri} = \frac{\big< \left| r \right| \big>}{\big< \left| i \right| \big>}.
\end{equation}
Where $r$ is the measured Stokes parameter when the pure $i$ Stokes
parameter is input with $i,~r = I,~Q,~U$ or $V$.
The symbols $<>$ represent the average over the rectangular
boxes marked on the synthetic Stokes images shown in Fig. \ref{fig:stoksynth}
and $||$ implies that we have considered absolute values.
The total integration time is assumed to be 1~s in the simulation.

In order to evaluate the effects purely due to seeing, the values of $C_{ri}$
are corrected for the polarimetric response of the modulation 
schemes \cite{stenflo84}. 
The response matrix of a given modulation scheme is derived using six distinct 
states of input polarization and the corresponding output Stokes parameters 
\cite{beck05,nagaraju07}, in the absence of seeing effects.

   \begin{figure}
   \includegraphics[width=0.79\textwidth,height=0.3\textheight]{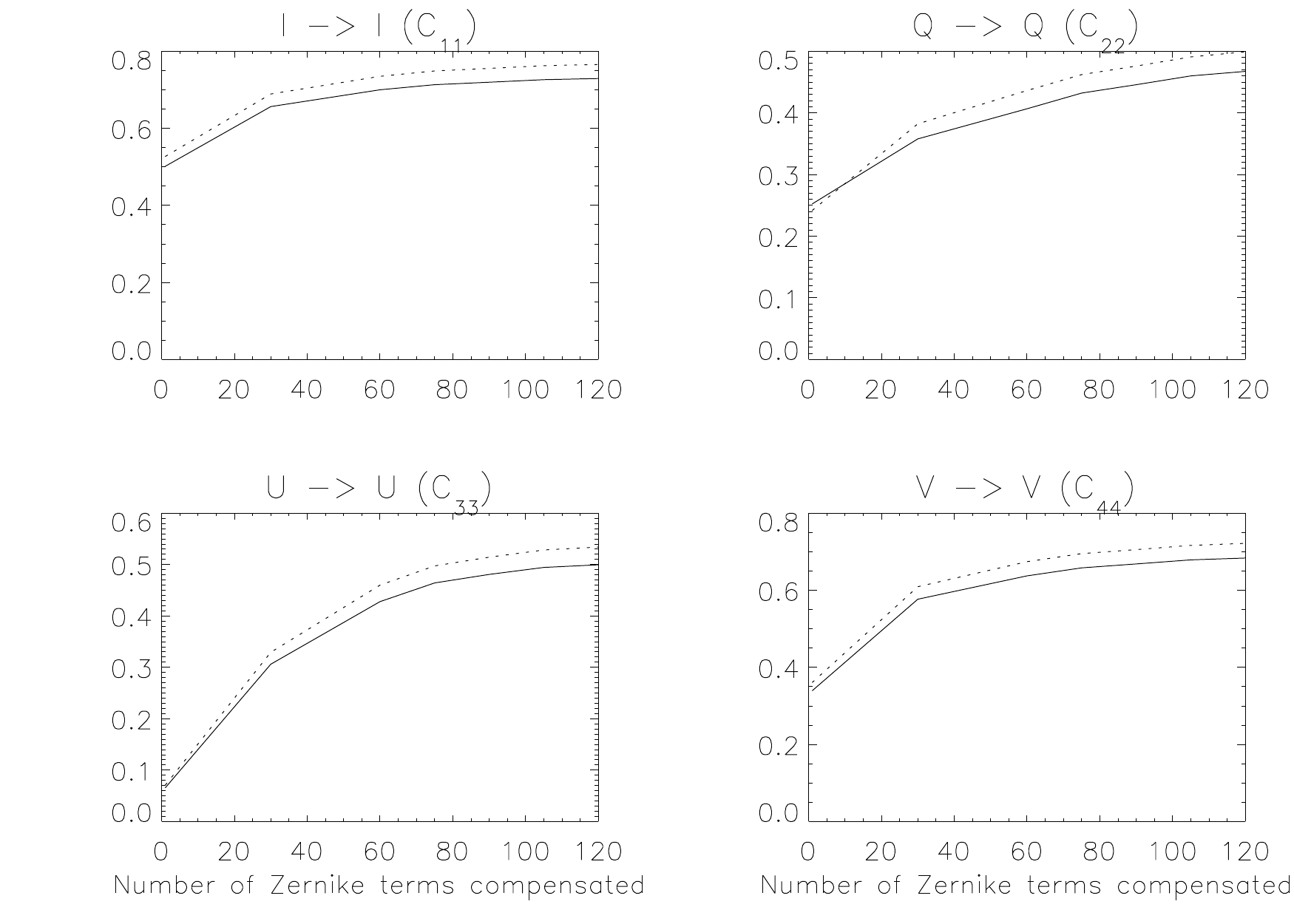}
   \caption{
   Plots of $C_{ii}$ as a function of number of lower order
   Zernike terms compensated for the modulation scheme 1 (solid curves) and
   the modulation scheme 2 (dotted curves). 
   The polarimeter is assumed to be in a single-beam setup.
   }
   \label{fig:eff}
   \end{figure}

Plots of the diagonal elements of $C_{ri}$ as a function of $J$ are shown in 
Fig. \ref{fig:eff}.
This figure suggests that the amplitude of the original Stokes signal is
more efficiently recovered with increase in $J$.
This is because of the reduction in image smearing with increase in $J$.
The values of $C_{ii}$ do not depend on the modulation frequency. 
Hence the plots are shown only for one modulation frequency case.
Further, the values of $C_{ii}$ depend neither on the polarimeter setup 
(single beam or dual beam) nor on the modulation scheme.

   \begin{figure}
   \includegraphics[width=0.99\textwidth,height=0.4\textheight]{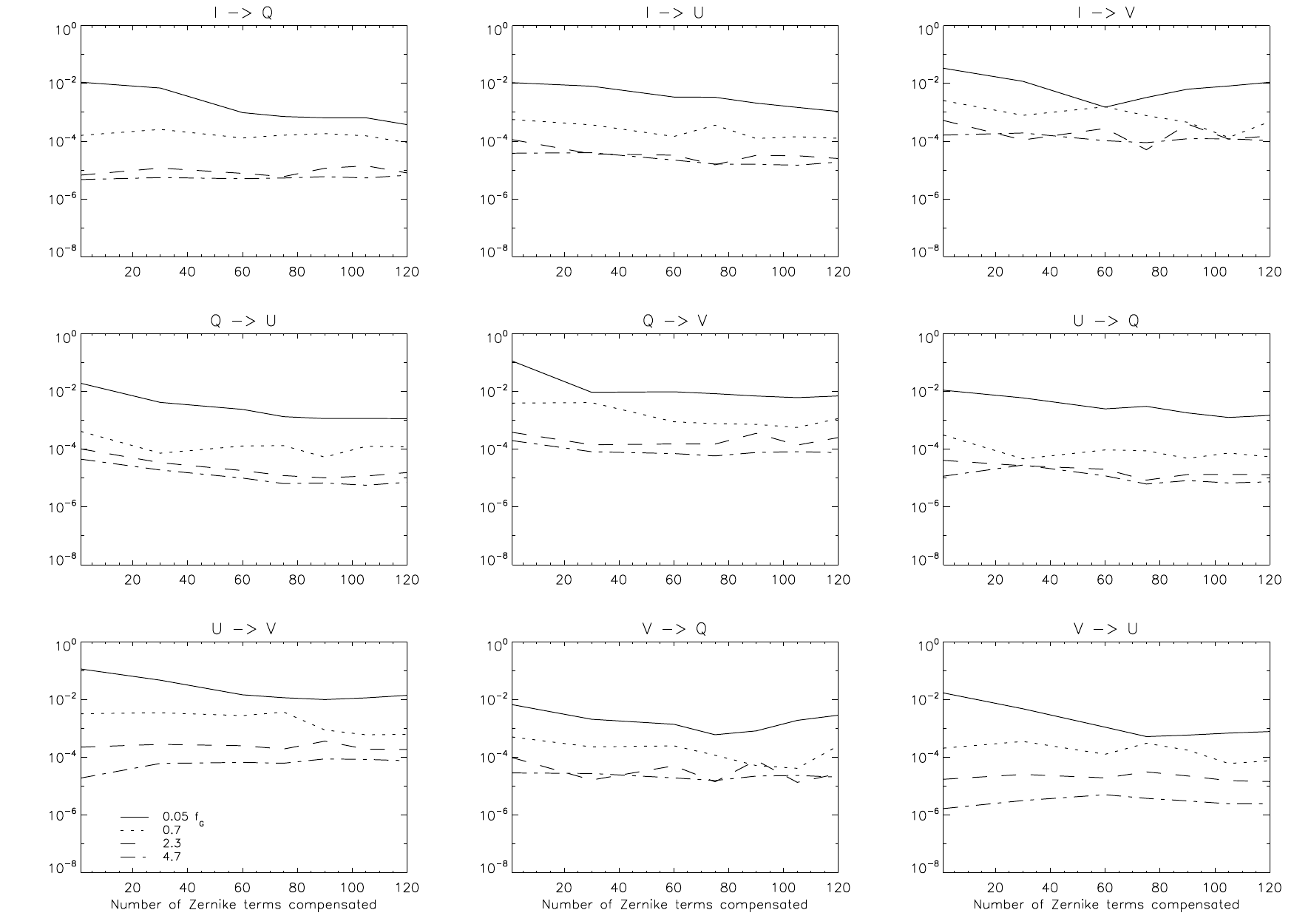}
   \caption{
   Plots of cross-talk elements in scheme~1 as a function of the number of 
   Zernike terms compensated and for different modulation frequencies.
   the polarimeter is assumed to be in a single-beam setup.
   }
   \label{fig:crosstalk_rotwp}
   \end{figure}
 
Plots of cross-talk among the Stokes parameters, i.e. the off-diagonal 
elements of $C_{ri}$, are shown in Fig. \ref{fig:crosstalk_rotwp}, 
as a function of $J$ and for different modulation frequencies of scheme 1. 
For brevity the error bars are not shown in the figure but we state here that 
the order of magnitude of 1-$\sigma$ statistical fluctuations of the cross-talk 
values range between $10^{-3}$ (towards slower modulation frequency) and 
$10^{-6}$ (towards higher modulation frequency). 
The modulation frequencies are expressed in Greenwood frequency 
\cite{greenwood77} which for the case of a single turbulent layer is given by 
\cite{hardy98}
\begin{equation}
f_G = 0.427 \frac{v}{r_0}
\end{equation}
For the simulation parameters considered in this paper the Greenwood frequency 
is 42.7 Hz.

In general, the cross-talk elements do not change significantly with increase 
in $J$. Only in a few cases at low modulation frequency
they show a slight decrease with $J$, maximum decrease is by an order of 
magnitude.
In contrast, the cross-talk reduction with increasing modulation frequency is 
much more significant. At a modulation frequency of 4.7$f_G$ the cross-talk
values are order $10^{-4}$ or less which are 2-3 orders
of magnitude smaller than at the simulated slowest modulation of 0.05$f_G$.

The result, that the cross-talk is not significantly affected by the number of 
aberration terms compensated, can be explained in terms of two effects which 
seem to balance each other.
With increase in number of lower order terms compensated the aberrations
in the images are reduced. This helps in improving the image quality and restoring
the Stokes intensity structures originally present.
This also helps in reducing the cross-talk values. But, with the improvement
in image quality the Stokes intensity gradients also increase. This effect
increases cross-talk values because of the presence of time varying 
residual higher order aberrations. As a result the cross-talk terms
remain more or less constant with $J$.

   \begin{figure}
   \includegraphics[width=0.99\textwidth,height=0.4\textheight]{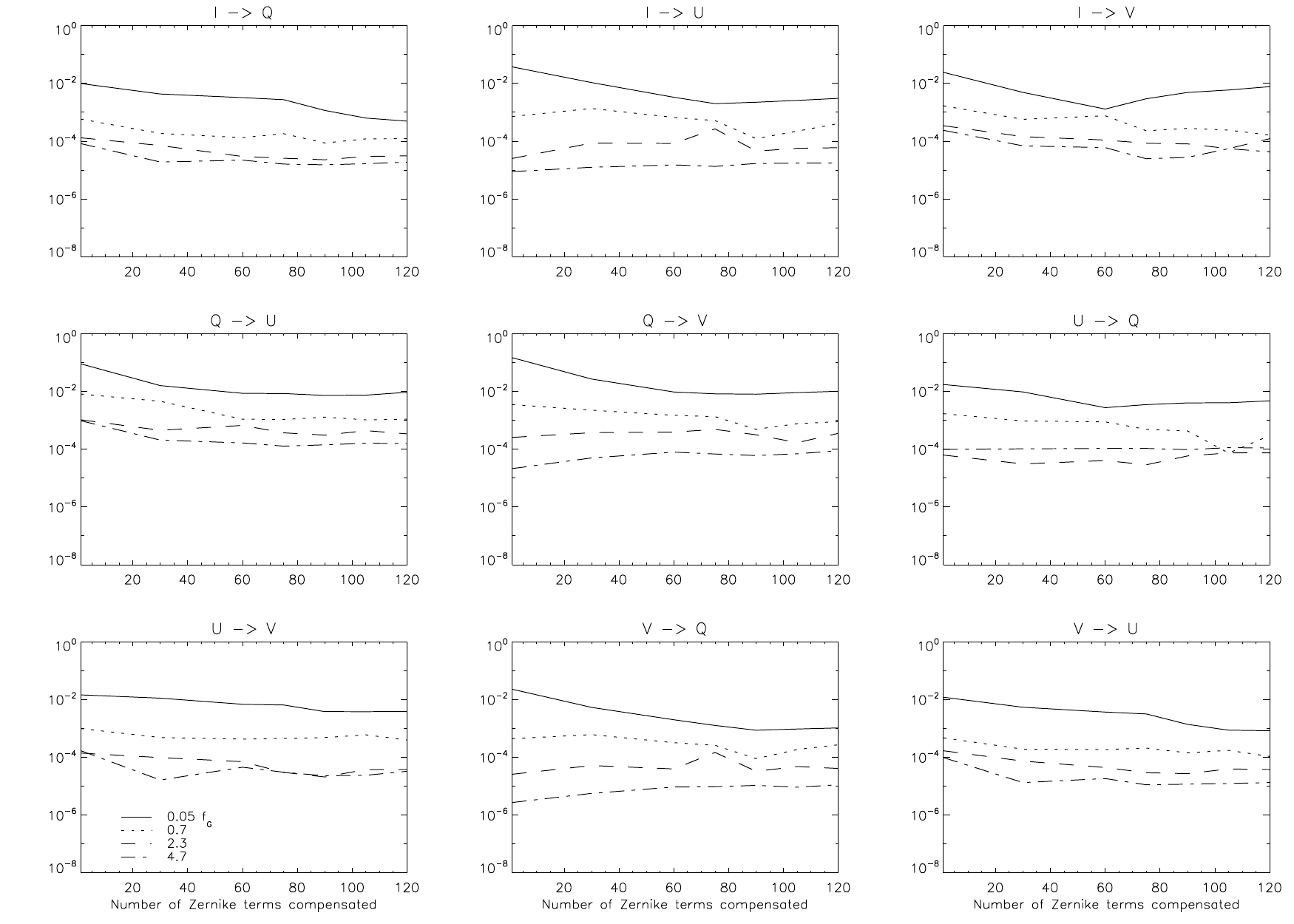}
   \caption{
   Same as Fig. \ref{fig:crosstalk_rotwp} but for scheme 2.
   }
   \label{fig:crosstalk_flc}
   \end{figure}

Results from scheme~2 also show the same trend 
(see Fig. \ref{fig:crosstalk_flc}). The cross-talk values are not different
from scheme~1 except in few cases. The statistical distribution of cross-talk 
values in scheme~2 are comparable with that of scheme 1. 
The comparison of cross-talk values between scheme~1 and scheme~2 suggests 
that choosing between discrete and continuous modulation schemes for 
polarization modulation is not so crucial as far as seeing induced cross-talk 
is concerned. 

   \begin{figure}
   \includegraphics[width=0.99\textwidth,height=0.4\textheight]{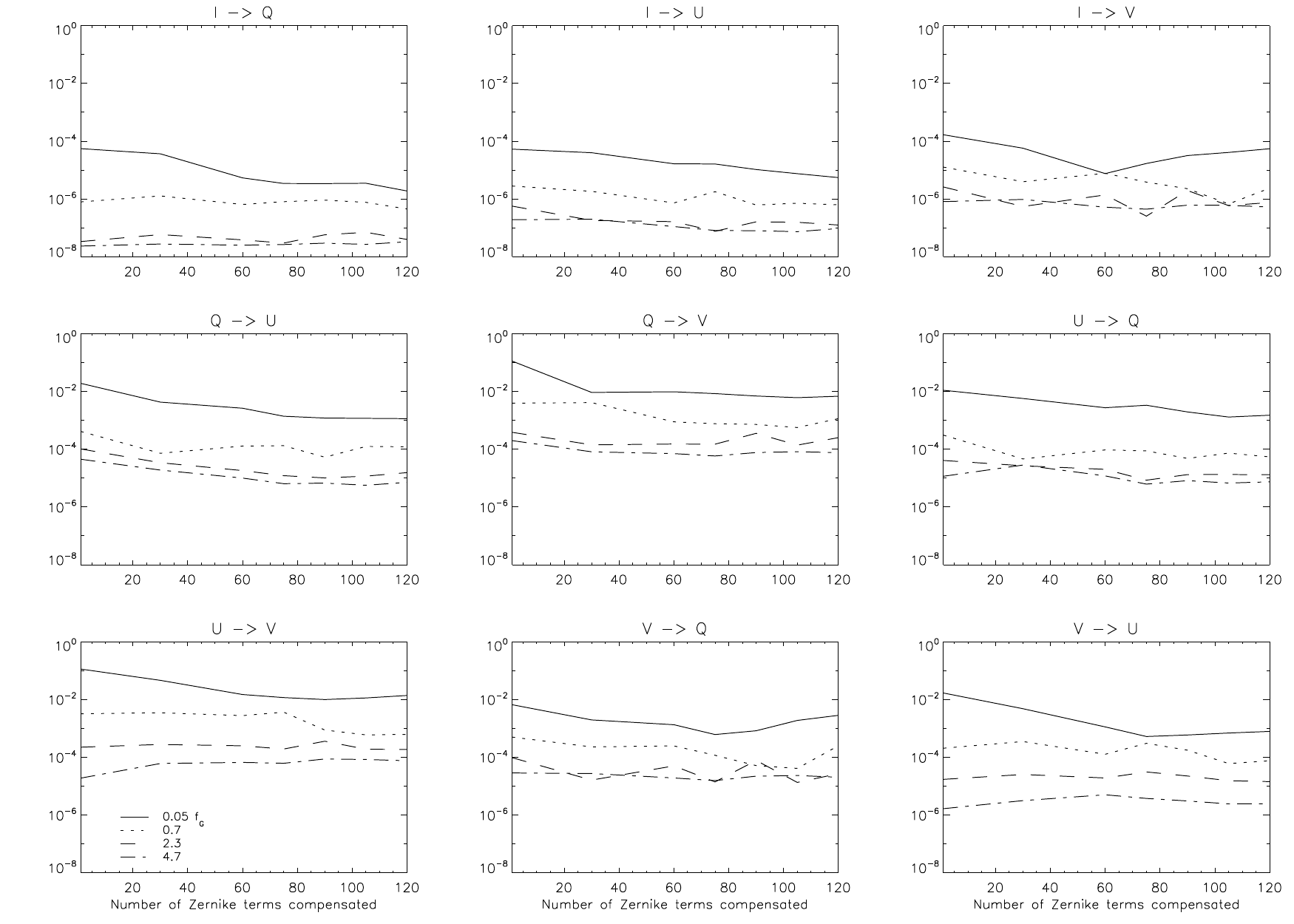}
   \caption{
   Same as Fig. \ref{fig:crosstalk_rotwp} but for dual-beam setup.
   }
   \label{fig:crosstalk_rotwp_2beam}
   \end{figure}

Plots of cross-talk evaluated in a dual-beam polarimeter setup are shown in 
Fig. \ref{fig:crosstalk_rotwp_2beam} for scheme 1. 
A beam imbalance of 1\% is assumed between the orthogonally polarized beams 
\cite{judge04}.
The statistical distribution of cross-talk values is comparable to the single 
beam setup except in the case of Stokes $I$ to $Q$, $U$ and $V$ cross-talk for 
which the range is between $10^{-5}$ and $10^{-8}$.
Fig. \ref{fig:crosstalk_rotwp_2beam} suggests the dual-beam setup helps in 
significant reduction of intensity to polarization cross-talk. 
The reduction in Stokes $I$ to $Q$, $U$ and $V$ cross-talk is about two orders 
of magnitude compared to the single-beam setup. 
On the other hand there is no change in the values of cross-talk among $Q$, $U$
and $V$. The reasons for reduction in cross-talk from $I$ to $Q$, $U$ and
$V$ in a dual-beam setup but no change in cross-talk among $Q$, $U$ and $V$ 
compared to a singlebeam setup are as follows. 
If $I^{\pm}$, $Q^{\pm}$, $U^{\pm}$ and $V^{\pm}$ represent the Stokes images 
obtained by demodulating the respective modulated intensities of the 
orthogonally polarized beams then the Stokes images of the combined beam are 
given by
\begin{eqnarray}
I = \frac{I^+ + I^-}{2}, \\\nonumber
Q = \frac{Q^+ - Q^-}{2}, \\\nonumber
U = \frac{U^+ - U^-}{2}, \\\nonumber
V = \frac{V^+ - V^-}{2}. \\\nonumber
\end{eqnarray}

Since the Stokes $Q$, $U$, and $V$ have opposite signs in the orthogonally 
polarized beams while Stokes $I$ has the same sign, cross-talk from $I$ 
to $Q$, $U$, and $V$ is cancelled out except for the gain difference whereas 
the cross-talk among $Q$, $U$, and $V$ will be just the average of 
the cross-talk present in the individual beams. 
As it was found for the single-beam case, there is not much difference in 
cross-talk values between scheme 1 and 2\ in the dual-beam setup also. 
Hence the plots of cross-talk values for scheme~2\ in the dual-beam setup are 
not shown. 


The simulations are carried out under the assumption of an ideal AO system.
Which means that the AO system fully compensates the $J$ number of lowest order
Zernike terms with infinite frequency bandwidth.
However in practice, a real AO has a finite band width and phase errors are
not eliminated completely. 
Because of the residual low order phase errors the image quality is
reduced which results in a smaller Stokes intensity gradient.
This effect tends to reduce cross-talk. 
However, because of the finite bandwidth of a real AO there 
high frequency low order phase errors still remain \cite{rimmele03} 
which tend towards increasing cross-talk.
This suggests that the amount of cross-talk in the case of a real AO will
depend on the amplitude of the residual low order phase errors within and
outside the AO correction bandwidth in addition to the uncorrected higher order
aberrations.

\section{Conclusion}
We have presented in this paper the analysis of seeing induced polarization
cross-talk through numerical simulation. We confirm in this work the 
earlier results by Lites \cite{lites87} and Judge et al. \cite{judge04} that
a considerable reduction in cross-talk can be achieved by increasing
the polarization modulation frequency. 
In addition we have shown that the seeing induced cross-talk remains
practically independent of the number of aberration terms compensated by an 
AO system. 
This is due to the presence of residual higher order image aberrations.
However, an AO system helps in improving the seeing affected
signal-to-noise ratio through increased spatial resolution.

\section*{Acknowledgments}
This project is carried out with financial support from the European Community
(project EST, number 212482).


\end{document}